\documentclass{aa}  

\usepackage{graphicx}

\usepackage{url}
\usepackage[breaklinks,colorlinks,citecolor=blue]{hyperref}
\usepackage{txfonts}
\begin{document}

   \title{IC~4499 revised: spectro-photometric evidence of small light-element variations}


   \author{E. Dalessandro\inst{1}
   \and
   C. Lardo\inst{2}
   \and
   M. Cadelano\inst{1,3}
   \and
   S. Saracino\inst{1,3}
   \and 
   N. Bastian\inst{4}
   \and
   A. Mucciarelli\inst{1,3}
   \and
  M. Salaris\inst{4}
  \and 
  P. Stetson\inst{5}
  \and 
  E. Pancino\inst{6,7}
}

   \institute{INAF -- Astrophysics and Space Science Observatory Bologna, Via Gobetti 93/3 I-40129 Bologna, Italy\\
   \email{emanuele.dalessandro@oabo.inaf.it}
   \and
   Laboratoire d' Astrophysique, Ecole Polytechnique F\'ed\'erale de Lausanne, Observatoire de Sauverny, CH-1290 Versoix, CH
   \and
Dipartimento di Fisica e Astronomia, Via Gobetti 93/2 I-40129 Bologna, Italy
  \and
   Astrophysics Research Institute, Liverpool John Moores University, 146 Brownlow Hill, Liverpool, L3 5RF, UK
   \and
   Herzberg Astronomy and Astrophysics, National Research Council Canada, 5071 West Saanich Road, Victoria, BC V9E 2E7, Canada
    \and
   INAF-Osservatorio Astrofisico di Arcetri, Largo Enrico Fermi 5, 50125, Firenze, Italy
   \and
   ASI Science Data Center, Via del Politecnico SNC, I-00133 Rome
 }

   \date{Received March XX, 2018; accepted XXX XX, 2018}

 
  \abstract{It has been suggested that IC~4499 is one of the very few old globulars 
to not host multiple populations with light-element variations. 
To follow-up on this very interesting result, here we make use of 
accurate HST photometry and FLAMES@VLT high-resolution spectroscopy to investigate in more detail the stellar population
properties of this system. 
We find that the red giant branch of the cluster 
is clearly bimodal in near-UV -- optical colour-magnitude diagrams, thus suggesting that IC~4499 is actually composed 
by two sub-populations of stars with different nitrogen abundances. This represents the first detection of multiple populations
in IC~4499.
Consistently, we also find that one star out of six is Na-rich to some extent, while we do not detect any evidence of
intrinsic spread in both Mg and O.

The number ratio between stars with normal and enriched nitrogen is in good agreement
with the number ratio - mass trend observed in Galactic globular clusters. Also, as typically found in other systems, 
nitrogen rich stars are more centrally concentrated than normal stars, although this result cannot be considered conclusive because
of the limited field of view covered by our observations ($\sim 1r_h$).

On the contrary, we observe that both the RGB UV color spread, which is a proxy of N variations, and Na abundance variations,
are significantly smaller than those observed in Milky Way globular clusters with mass and metallicity comparable to IC~4499. 
The modest N and Na spreads observed in this system can be tentatively connected 
to the fact that IC~4499 likely formed in a disrupted dwarf galaxy orbiting the Milky Way, as previously proposed based on its
orbit. 
}

   \keywords{globular clusters: general -- globular clusters: individual (IC~4499) --
-- stars: Hertzsprung-Russel and C-M diagrams -- stars: abundances -- techniques: photometric -- techniques:
spectroscopic  }

\titlerunning{MPs in IC~4499}
\authorrunning{Dalessandro et al.}
   \maketitle



\section{Introduction}
Globular clusters (GCs) exhibit intrinsic star-to-star variations in their light 
element content. In fact, while some GC stars have the same light-element abundances 
as the field at the same metallicity (first population/generation--FP), 
others show enhanced N and Na along with depleted C and O abundances 
(second population/generation--SP). The manifestation of such light-element 
inhomogeneities is referred  to as multiple populations (MPs).
A number of scenarios have been proposed over the years to explain 
the formation of MPs \citep[e.g.][]{decressin07,dercole08,bastian13,denissenkov14,dantona16,gieles18}, however, 
their origin is still strongly debated.

Nearly all massive and old Galactic GCs host MPs 
(e.g., \citealt{piotto15,milone17,bragaglia17}). In addition, MPs are also found 
in the Magellanic Clouds stellar clusters \citep{mucciarelli09,dalessandro16}, 
in GCs in dwarf galaxies such as Fornax \citep{larsen12,larsen18}, in the M31 GC system \citep{schiavon13} and there
are strong indications (though indirect) that they are ubiquitous in stellar clusters in massive elliptical galaxies 
(e.g., \citealt{chung11}).  

However, it is currently unclear which GC property drives the onset of MPs in GCs.
One property that has a strong effect on both the relative fraction of SP stars as well as the 
extent of the abundance spreads is 
the GC present-day mass (e.g., \citealt{carretta10,schiavon13,milone17}).  
Low mass GCs ($<10^5 M_{\odot}$) have $\sim50$\% fractions of SP stars  
increasing to $\sim90$\% in the most massive GCs ($\sim10^6 M_{\odot}$; \citealp{milone17}).  
By extrapolating this relation to the open cluster mass regime ($\sim 10^3-10^4 M_{\odot}$), 
we would expect to find $1-10$\% of SP, but they have not been found in these system.

Cluster age turns out to be another important parameter. Indeed, 
MPs have been found in intermediate age ($2-8$~Gyr) clusters in the Magellanic Clouds 
\citep{nied17,martocchia18}.
Conversely, massive ($\sim$ 10$^5$ M$_{\odot}$) clusters younger than $\sim$ 2 Gyr do not show any inhomogeneity 
in their light-element content or any photometric signatures that can be indicative of 
chemical anomalies \citep[e.g.][]{mucciarelli08,martocchia18}.

Only very few exceptions to this general emerging scheme are known in the literature.
One of them is the ancient ($t_{age}=12\pm0.75$Gyr;  
\citealt{dotter11}) GC IC~4499 \citep{walker11}, other classical cases are Ruprecth~106 \citep{villanova13} and E3
\citep{monaco18}.
IC~4499 has been reported to not host MPs based on 
ground based multi-band photometry by \citet{walker11}.
Given the age and mass ($M_V=-7.32$; \citealt{harris96} -
edition 2010) of this cluster, the lack of MPs is totally unexpected and extremely interesting.  
In fact, finding clusters that do not host MPs may lead to new insights into GC formation.

Here we use Hubble Space Telescope (HST) imaging targeting the cluster central regions and FLAMES@VLT high-resolution spectroscopy 
of six bright cluster members to conclusively establish whether IC~4499 hosts MPs or not.

The paper is structured as follows. In Section~2 the photometric observational database and analysis are presented. 
In Section 3 we detail on the spectroscopic analysis. In Section~4 the results obtained by
means of the photometric and spectroscopic data-sets are discussed. 
In Section~5 we summarise the most relevant results
and draw our conclusions.

\section{PHOTOMETRIC OBSERVATIONS AND DATA ANALYSIS}

\begin{figure}
	\includegraphics[width=\columnwidth]{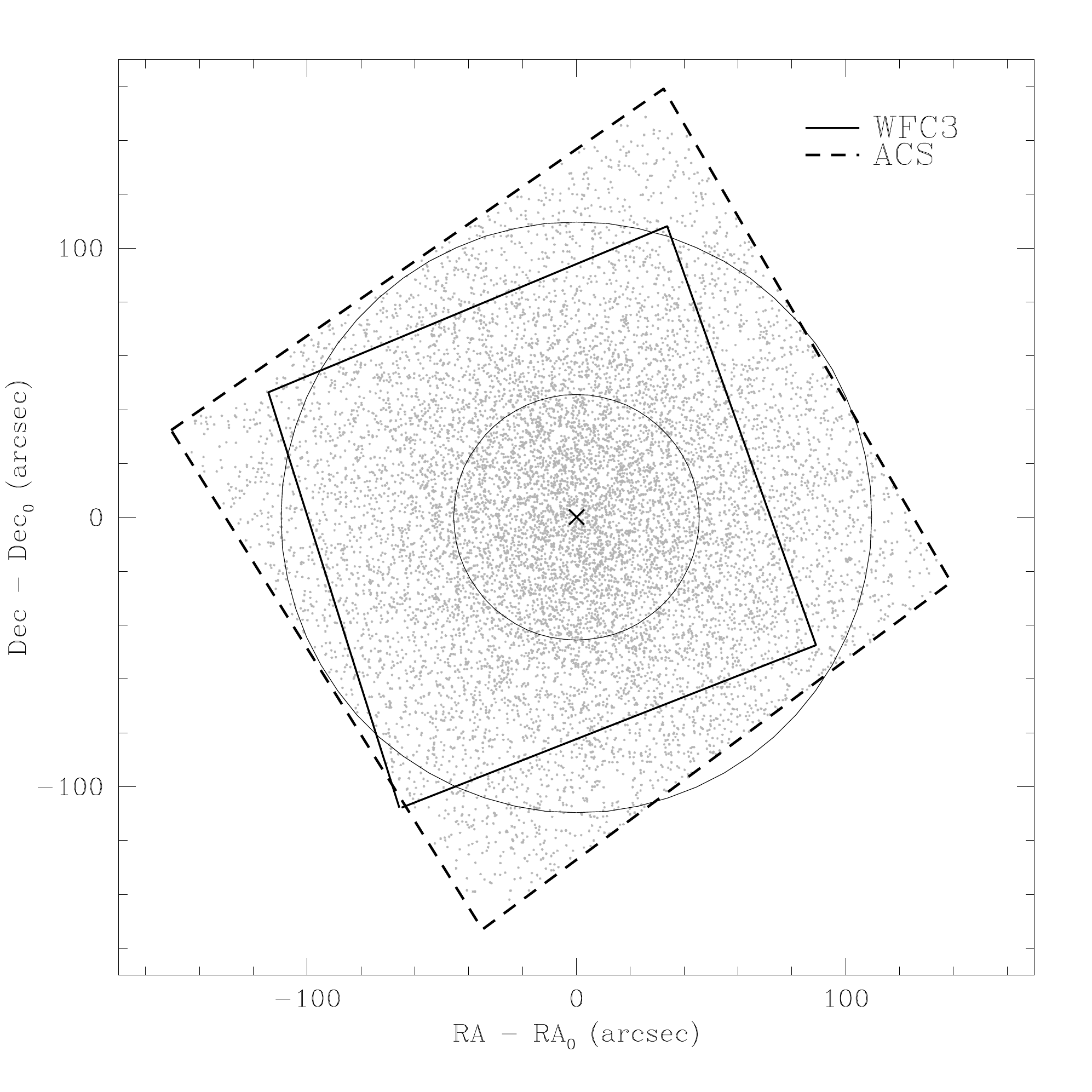}
    \caption{Map of the HST/WFC3 and HST/ACS images FOVs. The two black circles represent
    the cluster core and half mass radii from \citet{harris96}.}
    \label{fig:map}
\end{figure}

\begin{figure*}
\centering
\includegraphics[width=0.8\textwidth]{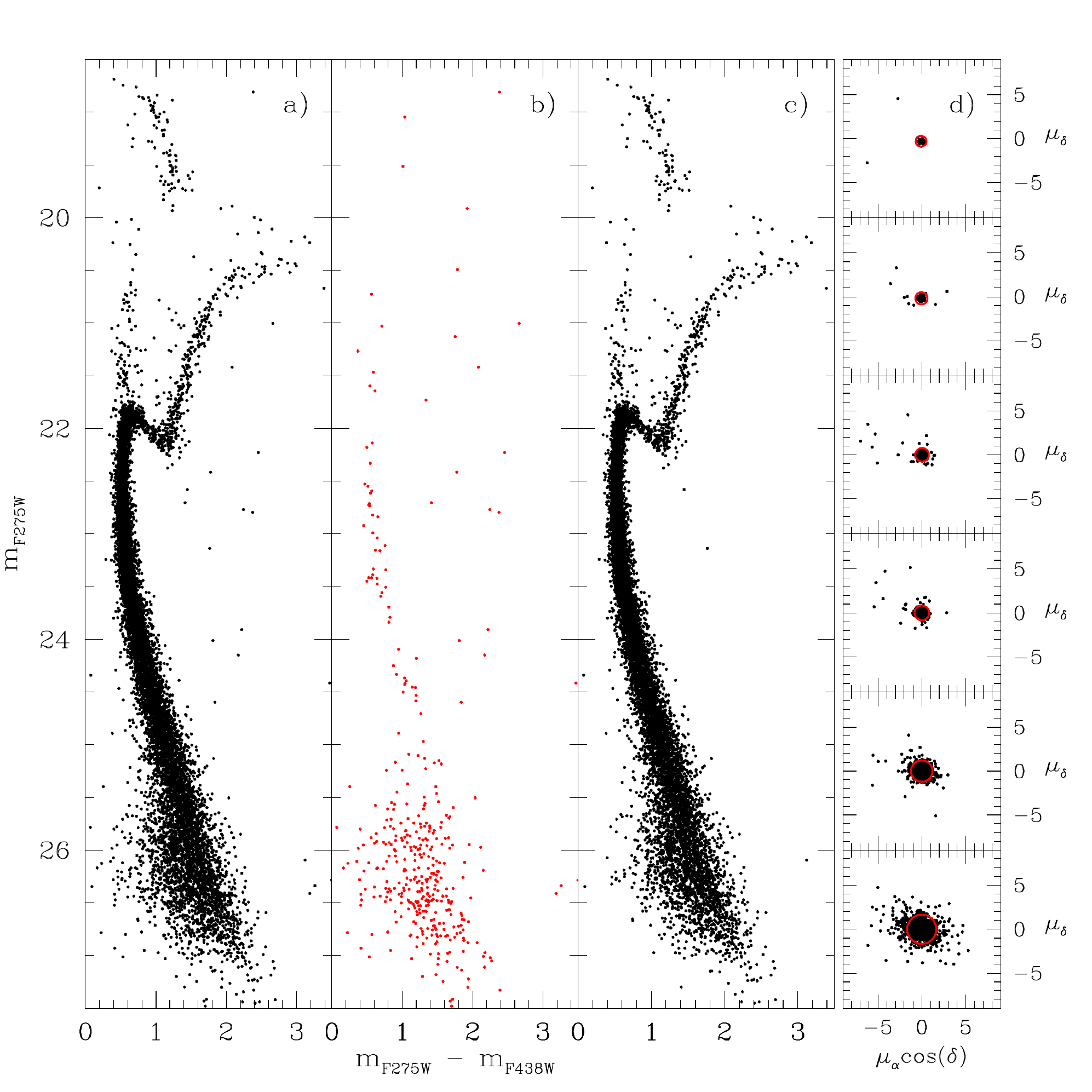}
\caption{{\it Panel a)}: ($m_{F275W}$, $m_{F275W}-m_{F438W}$) CMD of all the stars in common between the ACS and WFC3
catalogs; {\it Panel b)}: 
CMD of all the likely contaminating objects as selected from the VPDs; 
{\it Panel c)}: relative PM-cleaned CMD obtained by using only likely cluster members;
{\it Panels d)}: VPDs at different magnitude levels. The solid red circle represents the fiducial region used to select 
likely cluster members.}
\label{fig:ic4499PM}
\end{figure*}

\subsection{Photometric data and reduction}
This work is based on two HST data-sets.  
One is composed of near-UV and optical images obtained through the Wide Field Camera 3 (WFC3) 
UVIS channel (GO 14723, PI: Dalessandro). Eight images were acquired with the F275W filter and exposure time $t_{exp}=1100$ s, 
four F336W images with  $t_{exp}=650$s  and additional four exposures acquired with the F438W band and $t_{exp}=100$s.  
The other programme consists of five images acquired with the F606W filter (one with $t_{exp}=60$s and four with $t_{exp}=603$s)
and five F814W images (one with $t_{exp}=65$ s and four with $t_{exp}=636$s) obtained with 
the Wide Field Camera of the Advanced Camera for Surveys (GO: 11586, PI: Dotter). 
In both data-sets, an appropriate dither pattern of a few arcseconds has been adopted for each pointing 
in order to fill the inter-chip gaps and avoid spurious effects due to bad pixels. 
The field of view (FOV) covered by the two datasets is shown in Figure~\ref{fig:map}.

The photometric analysis has been initially performed independently for each data-set, camera and chip 
by using {\tt DAOPHOT IV} \citep{stetson87}. Tens of bright and isolated stars have been selected in each frame to 
model the point spread function (PSF), which has been eventually applied 
to all sources detected in each image above $5\sigma$, where $\sigma$ is the standard deviation of 
the background counts. We then created a master catalog composed of stars detected in 
at least half of the images for each available filter and two bands. 
At the corresponding positions of stars in this final master-list, a fit was forced with {\tt DAOPHOT/ALLFRAME} \citep{stetson94}
in each frame of the two data-sets. 
For each star thus recovered, multiple magnitude estimates obtained in each chip were homogenised by using 
{\tt DAOMATCH} and {\tt DAOMASTER}, and their weighted mean and standard deviation were finally adopted as 
star magnitude and photometric error.

Instrumental magnitudes of both the WFC3 and ACS catalogs were reported to the 
VEGAMAG photometric system. For the case of the WFC3, we used the equations and zero-points reported in the HST web page, 
while for the ACS we used the stars in common with the catalog used by \citet{dotter11} and 
publicly available in the ACS Survey of Galactic Globular Clusters 
page\footnote{\url{http://www.astro.ufl.edu/~ata/public_hstgc/databases.html}}.
The resulting ($m_{F275W}$, $m_{F275W}-m_{F438W}$) colour magnitude diagram (CMD) is shown in Figure~\ref{fig:ic4499PM}.

Instrumental coordinates were reported to the absolute coordinate system ($\alpha, \delta$)
by using the stars in common with the \citet{dotter11} catalog as secondary astrometric standards 
and by means of the software {\tt CataXcorr}\footnote{We used CataXcorr, a code which is specifically developed 
to perform accurate astrometric solutions. 
It has been developed by P. Montegriffo at INAF- Osservatorio Astronomico di Bologna. 
This package is available at http://davide2.bo.astro.it/$\sim$paolo/Main/CataPack.html, and has been successfully used in a large number 
of papers by our group in the past 10 years.}.

\subsection{Relative Proper Motion}\label{PM}
By taking advantage of the large temporal baseline existing between the WFC3 and ACS data-sets ($\Delta$T = 6.915 yr),
we performed a relative proper motion (PM) analysis to clean the cluster CMD from field interlopers.
We note that IC~4499 is at heliocentric distance $d=18.8$ Kpc thus the PM information coming from 
the Gaia Data Release 2 are available only for the brighter portion of the red giant branch and suffer 
from significant uncertainties.

For the proper motion analysis we adopted the approach described in \citet{dalessandro13} (see also
\citealt{massari13,bellini14,cadelano17}).
The procedure consists in measuring the instrumental position displacements 
of the stars detected in both epochs, 
once a common distortion-free reference frame is defined. 
The reference frame adopted in this analysis is the geometric 
distortion corrected ACS catalog (see Section 2.1). 

For each data-set we derived mean instrumental positions (x,y) as the average of the positions of stars detected 
in half of the entire number of images (see Section~2). In the WFC3 case (x,y) have been corrected for geometric distortions  
by applying the equations published 
in \citet{bb09} for the $F275W$ UVIS filter. For the ACS catalog we adopted the ACS/WFC Distortion Correction Tables 
({\textrm IDCTAB}) provided on the dedicated page of the Space Telescope Science Institute. 

We then estimated accurate transformations between the WFC3 catalog and the reference frame. 
To this aim we selected a sample of $\sim$ 2000 
stars having magnitude 18.0 $\leq$ $m_{F814W}$ $\leq$ 20.5 (corresponding to magnitudes 19.5 $\leq$ $m_{F336W}$ $\leq$ 22.0), 
which are likely cluster members on the basis of their positions in the CMD 
(stars distributed along the lower RGB, the sub-giant branch and the upper main sequence). 
We then applied a six-parameter linear transformation between the two epochs, 
treating each chip of the WFC3 separately, in order to maximise the accuracy.

The derived transformations, have been then applied to 
all the stars in common between the two catalogs. 
The relative PM is finally determined by measuring 
the difference of the mean (x,y) positions of each star in the two epochs, 
divided by their temporal baseline. 
Such displacements are in units of pixels yr$^{-1}$. We finally converted the PMs into absolute units (mas yr$^{-1}$) 
by multiplying the measured displacements by the pixel scale of the master frame (0.05''/pixel). 
 
Results of the PM analysis are shown in Figure \ref{fig:ic4499PM}. 
In Panel a), the ($m_{F275W}, m_{F275W}-m_{F438W}$) CMD of the stars in common between ACS and WFC3 is shown and 
the relative PM measurements are shown as vector point diagrams (VPD) in Panels d) 
for 6 different bins of 1.5 mag each. 
A $\sigma$-clipping procedure was applied to separate field stars from cluster members in each VPD. 
Stars located within the red circles (likely cluster members) are
plotted as black points in Panel (c) of Figure \ref{fig:ic4499PM}, while those located outside the circles (likely field interlopers) are shown 
as red points in the Panel b).

\begin{figure}
\includegraphics[width=\columnwidth]{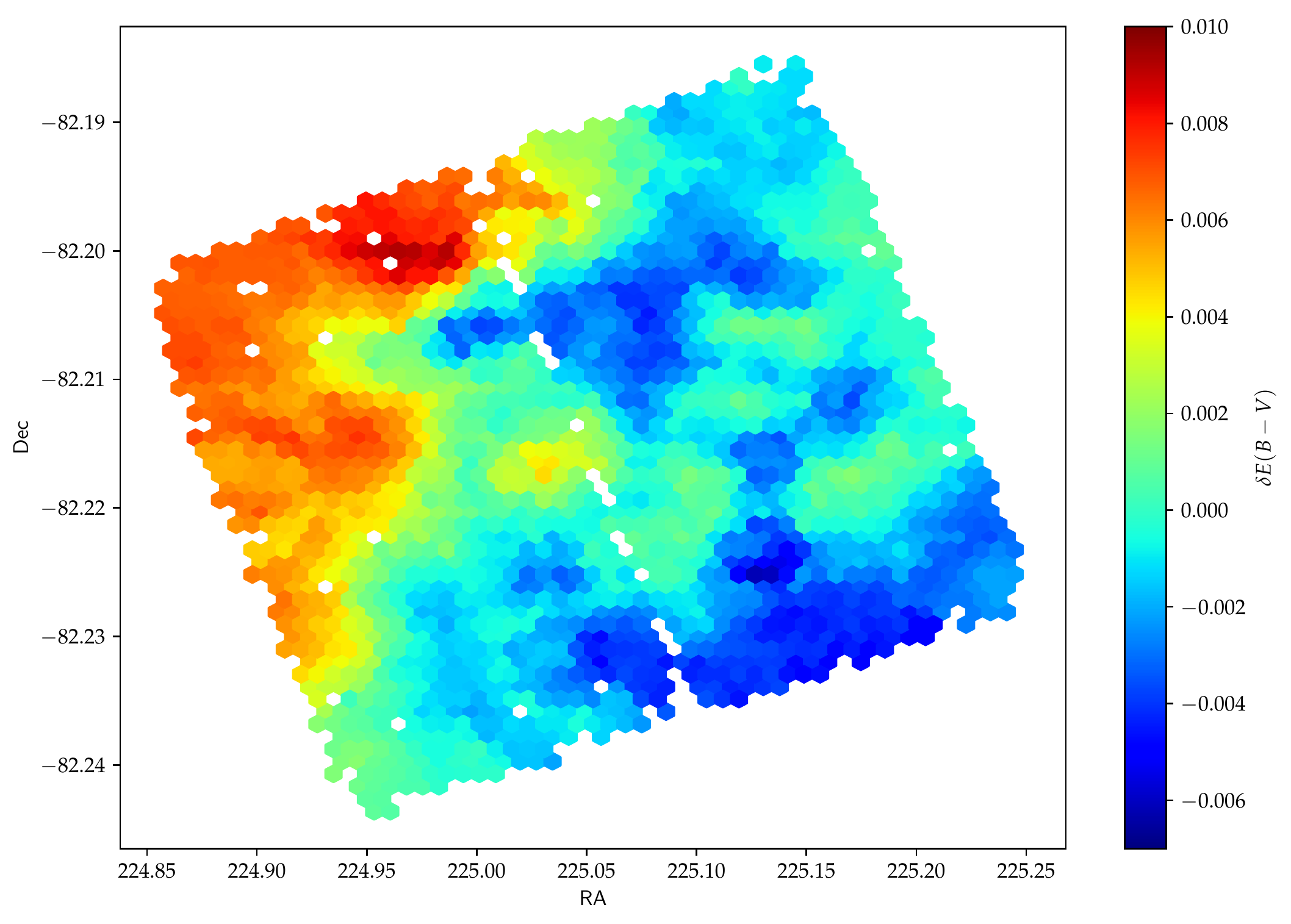}
	\caption{Differential reddening map of IC 4499 in the ACS+WFC3 FOV obtained as described in Section~2.3.
	\label{fig:ic4499_dr} }
\end{figure}  

\subsection{Differential reddening}
IC~4499 has a moderately large color excess E(B-V) $\approx$ 0.23 \citep{harris96}, 
hence for the goal of this work it is important to quantify and eventually correct
for any reddening variations within the HST field of view.

\begin{figure}
\includegraphics[width=\columnwidth]{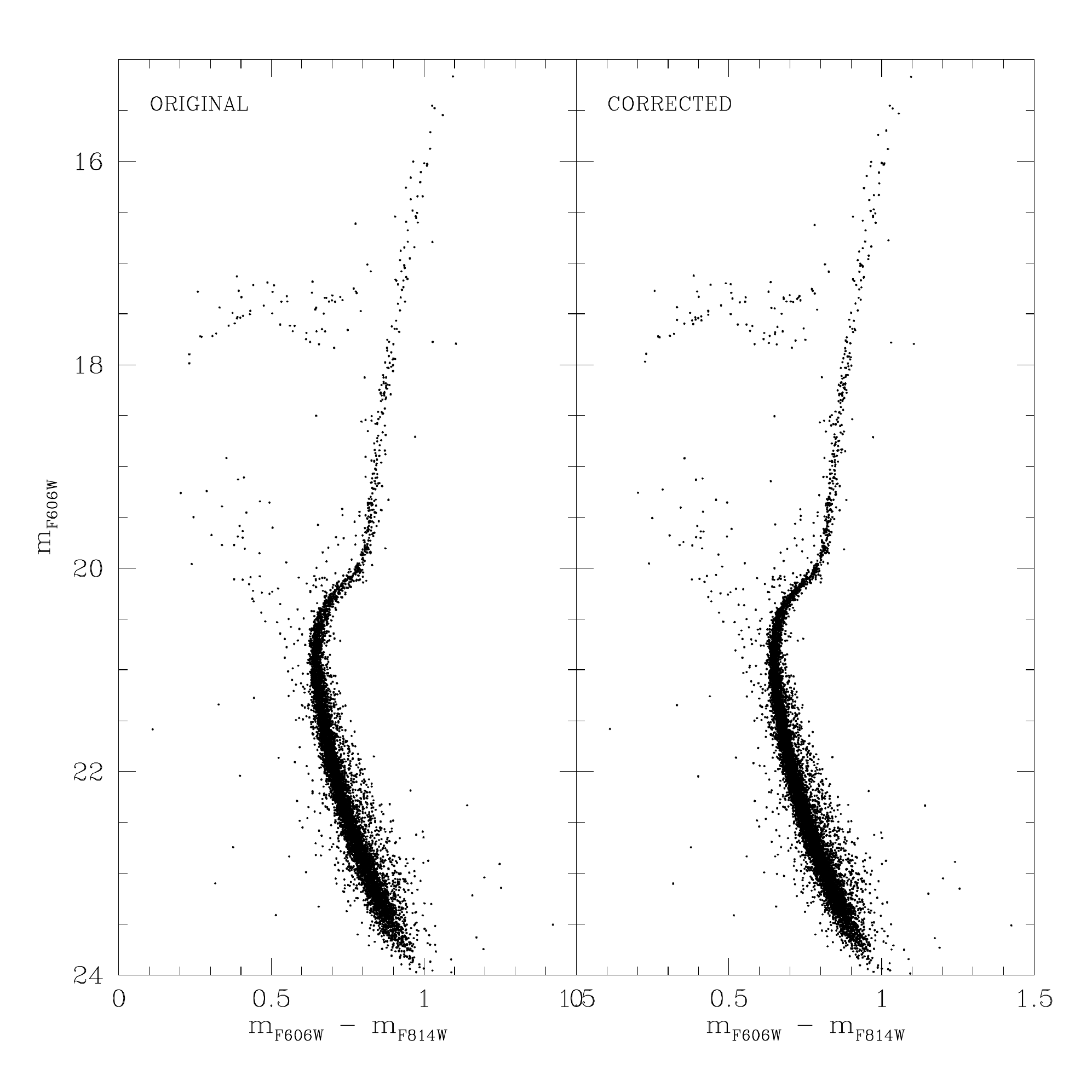}
	\caption{PM-cleaned ($m_{F606W}, m_{F606W}-m_{F814W}$) CMDs of IC4499 
	 before (left panel) and after (right panel) differential reddening correction. \label{fig:ic4499_comp} }
\end{figure}

Differential reddening was estimated 
by using likely member stars (as constrained from relative PMs analysis described in Section~\ref{PM})
selected in the magnitude range 18.5 $\leq$ $m_{F606W}$ $\leq$ 21.0. For these stars  
a mean ridge-line (MRL) in the ($m_{F606W}, m_{F606W}-m_{F814W}$) CMD (Figure~\ref{fig:ic4499_dr}) was defined. 
We then computed the geometric distance ($\Delta X$) to the MRL
for each star within $3\sigma$ from it, where $\sigma$ is the colour 
spread around the MRL.
By definition the MRL has $\Delta X=0$ and all stars are distributed around
this value.
Then for each star in the catalog, differential reddening is estimated by computing the mean of the $\Delta X$ values 
of the 20 nearest (in space) selected stars. We have verified that in this specific case 
this number of neighbour stars is the best compromise between good statistics and spatial resolution. 
Results do not change significantly if a slightly different number of neighbour stars (from 10 to 30) is adopted. 
Finally, $\Delta X$ values are transformed into 
differential reddening $\delta E(B-V)$ by using the following equation: 

\begin{equation}
\delta E(B-V) = \frac{\Delta X}{\sqrt{2R^{2}_{F606W}+R^{2}_{F814W}-2R_{F606W}R_{F814W}}}
\end{equation}

where $R_{F606W}$ and $R_{F814W}$ are the exctintion coefficients in the two bands respectively.
We adopted $R_{F606W}$ = 2.874 and $R_{F814W}$ = 1.884 as given by \citet{casagrande14}.  
The resulting reddening distribution within the WFC3-ACS FOV is shown in Figure~\ref{fig:ic4499_dr}. 
A maximum $\delta E(B-V)\sim0.015$ mag is found within the HST FOV. 
Figure \ref{fig:ic4499_comp} 
shows the ($m_{F606W}, m_{F606W}-m_{F814W}$) CMD of IC~4499 before (left panel) 
and after the differential reddening corrections have been applied (right panel). 

In the following analysis we will use PM cleaned and differential reddening corrected CMDs.

\begin{table*}
\begin{footnotesize}
\begin{center}

\setlength{\tabcolsep}{0.2cm}

\caption{Coordinates, $V$ magnitude, radial velocities, and atmospheric parameters of IC 4499 stars observed with UVES.}
\label{PHOT_PAR}
\begin{tabular}{@{}rccccccc}
\hline
ID	  &  RA(J2000)  &   Dec(J2000)   &     V    &  v$_{\rm {r}}$             &     T$_{\rm {eff}} $  &  log g             & $\xi$t  \\
           &   (deg)         &    (deg)            &      (mag)   &   (km/s)             &     (K)           &  (dex)             &    (km/s) \\
          \hline
S354   &  224.4198320  &    -82.1760627 &   15.189 &   35.4  $\pm$	 0.4  &  4280   & 0.85  &  1.8  \\
S79    &  225.0212891  &    -82.1874560 &  15.230 &     38.6  $\pm$	 0.4  &  4300   & 0.90  &  1.8  \\
S634   &  224.9185887  &    -82.2678076 &  15.267 &    37.1  $\pm$	 0.4  &  4350   & 0.70  &  2.1  \\
S81    &  225.0442379  &    -82.1880345 &    15.588 &     37.0  $\pm$	 0.4  &  4430   & 0.90  &  1.7  \\
S639   &  224.9446419  &    -82.2710076 &   15.683 &     36.2  $\pm$	 0.4  &  4400   & 0.75  &  1.6  \\
S664   &  225.4816207  &    -82.1793317 &  15.822 &    37.7  $\pm$	 0.5  &  4600   & 1.10  &  1.9  \\

\hline

\end{tabular}

\end{center}
\end{footnotesize}
\end{table*}

\begin{figure*}
\includegraphics[width=\textwidth]{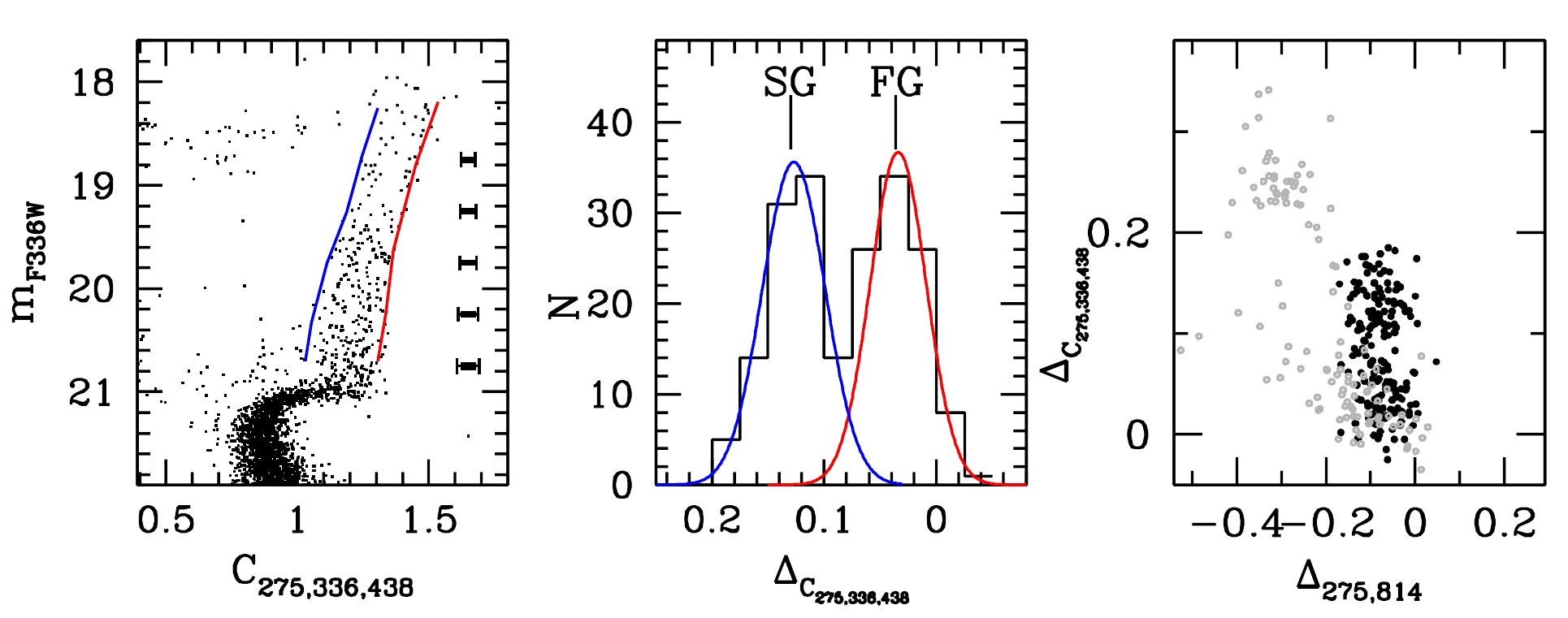}
	\caption{{\it Left panel}: (m$_{F 336W}$ , C$_{275,336,438}$) pseudo-colour diagram of IC~4499. Typical photometric errors
	for different magnitude bins are also shown. 
	{\it Middle panel:} Verticalised pseudo-color distribution $\Delta$C$_{275,336,438}$ of RGB stars in the 
	magnitude range $18.2< m_{F 336W}<20.5$.Two main components can be clearly identified 
	(SP and FP, in blue and red; respectively). Right panel: ($\Delta _{275,814}$, $\Delta$C$_{275,336,438}$) 
	colour-colour diagram (``chromosome map'') of IC~4499. Overplotted for reference is shown also 
	 the chromosome map of NGC~288 (light grey).
	 \label{fig:cmap}}
\end{figure*}

\section{SPECTROSCOPIC OBSERVATIONS AND DATA ANALYSIS}
\subsection{Observations and data reduction}
We used FLAMES@VLT 
observations in the combined MEDUSA+UVES mode (097.D-0111(A) - PI: Bastian), which allows
the simultaneous allocation of 8 UVES high-resolution 
and 132 MEDUSA mid-resolution
fibers. Only UVES spectra will be used in this work, 
while the MEDUSA ones will be presented in a forthcoming paper (Carmela Lardo et al. 2018, in preparation) 
devoted to study the metallicity and $\alpha$-element abundance distributions of the IC~4499 stars. 
UVES spectra were obtained through the 580 Red Arm spectral configuration, 
providing a spectral resolution of
R$\sim$47\ 000 [$\lambda$/$\Delta$ $\lambda$] in the wavelength range between 4800-6800 \AA.
A total of three exposures of 50 min each for the same targets configuration
has been secured in Service Mode during July 2016.
Targets have been selected along
the brightest portion of the RGB (V$\leq$16 mag) from the $UBVI$ photometric
catalog by \citet{stetson94}. Two out of
the eight available UVES fibers have been used to sample
the sky background and perform a robust sky subtraction
for each individual exposure. The basic parameters of 
the six targets are listed in
Table~\ref{PHOT_PAR}, along with other useful information.
Data reduction was performed by following the standard recipes provided by the latest version 
of the ESO UVES pipeline. 
UVES data 
were bias subtracted, flat fielded, wavelength calibrated, and extracted using the standard UVES/esorex routines. 
Spectra were corrected for telluric absorption using a synthetic spectrum as a template.
The typical signal-to-noise ratio (SNR) per pixel of spectra 
is $\sim$20 at $\sim$5300~\AA~and $\sim$35 at $\sim$6300~\AA.

\subsection{Radial velocity and equivalent width measurements}

Radial velocities and equivalent widths (EWs) were obtained with {\tt DAOSPEC}
\citep{daospec}, run through the interface {\tt 4DAO}\footnote{\url{http://www.cosmic-lab.eu/4dao/4dao.php}} \citep{mucciarelli13c}. 
Atomic lines used in the analysis were selected from the Vienna Atomic Database (VALD, \citealp{kupka99}).

{\tt DAOSPEC} \citep{daospec} automatically performs the line profile fitting adopting a saturated Gaussian
function. The FWHM for each spectrum has been estimated iteratively, using the wrapper {\tt 4DAO}, 
which automatically configures many of {\tt DAOSPEC} parameters and provides graphical tools 
to explore the results. Firstly, we performed an analysis using as input value the nominal FWHM of the grating, 
leaving {\tt DAOSPEC} free to re-adjust the FWHM in order to minimise the median 
value of the residuals distribution. 
The new value of the FWHM is used by {\tt 4DAO} as input value for a new run, 
until a convergence of 0.3 pixel is reached.

The two UVES arms were analysed separately and the agreement between the 
two velocity estimates is good
($\langle$ v$_{\rm r}^{\rm Low}$ -- v$_{\rm r}^{\rm Up}$  $\rangle$= 0.4, $\sigma$=0.4 km/s).
Measurements were based on $\simeq$ 350 absorption lines of different elements.
The resulting velocities and errors are listed in Table~\ref{PHOT_PAR}.
We measured an average  v$_{\rm r}$ = 37.0 $\pm$ 0.4 km/s  ($\sigma$= 1.1 km/s), that agrees with the value 
(v$_{\rm r}$ = 31.5 $\pm$ 0.4 km/s) measured by \citet{hankey11} from their lower resolution spectra (R$\sim$10000).
All six stars result to be member based on their radial velocities.

\subsection{Abundance analysis}\label{ABBONDANZE}
 
Chemical abundances for iron and light-elements (O, Na, and Mg) were derived using the {\tt GALA} \citep{gala} package 
and the classical EW method for abundance analysis.

Model atmospheres were calculated with the {\tt ATLAS9} code
assuming local thermodynamic equilibrium (LTE) and one-dimensional, plane-parallel geometry;
starting from $\alpha$-enhanced grid of models available in F. Castelli's website \citep{castelli03}.
For all the models we adopted an input metallicity of [A/H]=--1.5 dex, according to the value 
derived by \citet{hankey11}.

Initial atmospheric parameters were obtained from the \citet{stetson94} optical photometry.
First, T$_{\rm eff}$ was derived from the B--V colour using \citet{alonso99} and the reddening 
E(B --V) = 0.23. Surface gravities ($\log$(g)) were obtained assuming a stellar mass 
of 0.8 M$_{\odot}$ and a distance modulus of (m--M)$_{\rm V}$=17.08  (\citealp{harris96}; 2010 edition).
The bolometric correction (BC) was derived by adopting the relation BC--T$_{\rm eff}$  from \citet{alonso99}. 
Finally, the initial micro-turbulence velocity ($\xi$t) was set to 2 km/s for all stars. 
These atmospheric parameters were considered as initial estimates and were refined during the abundance analysis. 
T$_{\rm eff}$ is adjusted until there is no trend between the abundance from Fe I lines and the excitation potential. 
The surface gravity is optimised in order to minimise the difference between the abundance derived from neutral and 
single ionised iron. Finally, $\xi$t is determined by erasing any trend between the abundance from Fe I and the reduced EW.

The mean abundances of Table~\ref{ABUNDANCES} are
computed by averaging the abundances of the surviving lines weighted by 
the uncertainty on the abundance as obtained from the EW error. 
The random errors ($\sigma _{\rm rand}$ in Table~\ref{ABUNDANCES}) are computed as the dispersion of the mean normalised 
to the root mean square of the number of used lines.
For those elements for which only one line is available, the error associated with the 
individual abundance estimate is taken as the internal error. 

The errors due to the uncertainty in the atmospheric parameters are obtained varying atmospheric parameters one at the time, 
by $\Delta$T$_{\rm eff}$=$\pm$70K, $\Delta$log(g)=$\pm$0.2 dex, $\Delta$ $\xi$t=$\pm$0.5 kms$^{-1}$, and $\Delta$[M/H]=$\pm$0.1dex and 
redetermining abundances.
Results are shown in Table~\ref{ERRORS} for the star S354, assumed to represent the entire sample.
 The total error ($\sigma _{\rm tot}$), listed in Table~\ref{ABUNDANCES} is obtained by adding the errors in
quadrature. Fe abundances were measured for all stars in our sample, whereas O and Na abundances are available only for four stars.
Five stars have measured [Mg/Fe] abundance ratios.

\begin{table*}
\begin{footnotesize}
\begin{center}

\setlength{\tabcolsep}{0.18cm}

\caption{Abundance Ratios for the Observed Stars}
\label{ABUNDANCES}
\begin{tabular}{@{}rcccccccccccccccc}
\hline

ID    &    [FeI/H]   &  $\sigma$$_{\rm rand}$   &  $\sigma$$_{\rm tot}$    &  [FeII/H]   &  $\sigma$$_{\rm rand}$   &  $\sigma$$_{\rm tot}$    &  [O/Fe]   &  $\sigma$$_{\rm rand}$   &  $\sigma$$_{\rm tot}$    &  [Na/Fe]   &  $\sigma$$_{\rm rand}$   &  $\sigma$$_{\rm tot}$    &   $\Delta$NLTE &  [Mg/Fe]   &  $\sigma$$_{\rm rand}$   &  $\sigma$$_{\rm tot}$  \\

\hline

S354	&   --1.64 &  0.01  &   0.07  &	 --1.57  &  0.04 &   0.09 &  0.45  &   0.04 & 0.09  &  --0.09 &  0.03 &  0.09  &--0.07   & 0.40 & 0.04 &  0.09  \\
S79	&   --1.66 &  0.01  &   0.07  &	 --1.59  &  0.04 &   0.09 &  0.40  &   0.03 & 0.09  &  --0.10 &  0.04 &  0.09  &--0.07   & 0.37 & 0.04 &  0.09  \\
S634	&   --1.70 &  0.01  &   0.08  &	 --1.62  &  0.02 &   0.08 &  0.34  &   0.05 & 0.08  &    0.23 &  0.03 &  0.08  &--0.10   & 0.27 & 0.04 &  0.08  \\
S81	&   --1.69 &  0.01  &   0.06  &	 --1.61  &  0.05 &   0.10 &  0.54  &   0.05 & 0.10  &  --0.22 &  0.05 &  0.10  &--0.05   & 0.22 & 0.05 &  0.09  \\
S639	&   --1.56 &  0.01  &   0.07  &	 --1.52  &  0.05 &   0.07 &  0.33  &   0.03 & 0.07  &  --0.21 &  0.04 &  0.07  &--0.07   & 0.30 & 0.03 &  0.07  \\
S664	&   --1.61 &  0.01  &   0.09  &	 --1.52  &  0.06 &   0.10 &  0.43  &   0.05 & 0.10  &  --0.24 &  0.05 &  0.10  &--0.06   & 0.25 & 0.08 &  0.09  \\

\hline

\end{tabular}

\end{center}
\end{footnotesize}
\end{table*}

Corrections for departures from the LTE approximation are applied only for 
the Na lines, by interpolating on the grid of corrections calculated by 
\citet{lind11}. In general these corrections are of the 
order of $\Delta$ NaI$_{\rm{LTE-NLTE}}$= --0.07 dex.
Reference solar abundances are by \citet{grevesse98} except for oxygen, 
for which we adopted the value provided \citet{caffau08}.

\begin{table*}
\begin{footnotesize}
\begin{center}

\setlength{\tabcolsep}{0.25cm}

\caption{Estimated Errors on Abundances Due to Errors on Atmospheric Parameters for star S354.}
\label{ERRORS}
\begin{tabular}{@{}rcccccccc}
\hline
Element   &	+ $\Delta$T$_{\rm eff}$  &  + $\Delta$T$_{\rm eff}$   &  + $\Delta$log(g) &  + $\Delta$log(g)   &  + $\Delta \xi$t   &    + $\Delta \xi$t   &  +$\Delta$[M/H]   & -$\Delta$[M/H]\\
 \hline
Fe I    &	 0.09  &  -0.08   &  0.01   &  0.00   &-0.10   &  0.16  &  -0.00   &  0.00 \\
Fe II   &	-0.06  &   0.06   &  0.09   & -0.09    &-0.05   &  0.08  &   0.03   & -0.03 \\
O I     &	-0.01  &   0.01   &  0.09   & -0.09    &-0.02   &  0.02  &   0.04   & -0.04 \\
Na I    &	 0.06  &  -0.07   & -0.02   &  0.02    &-0.04   &  0.05  &  -0.01   &  0.01 \\
Mg I    &	 0.06  &  -0.06   & -0.02   &  0.02    &-0.10   &  0.14  &  -0.01   &  0.01 \\

\hline

\end{tabular}

\end{center}
\end{footnotesize}
\end{table*}

\begin{figure}
\includegraphics[width=\columnwidth]{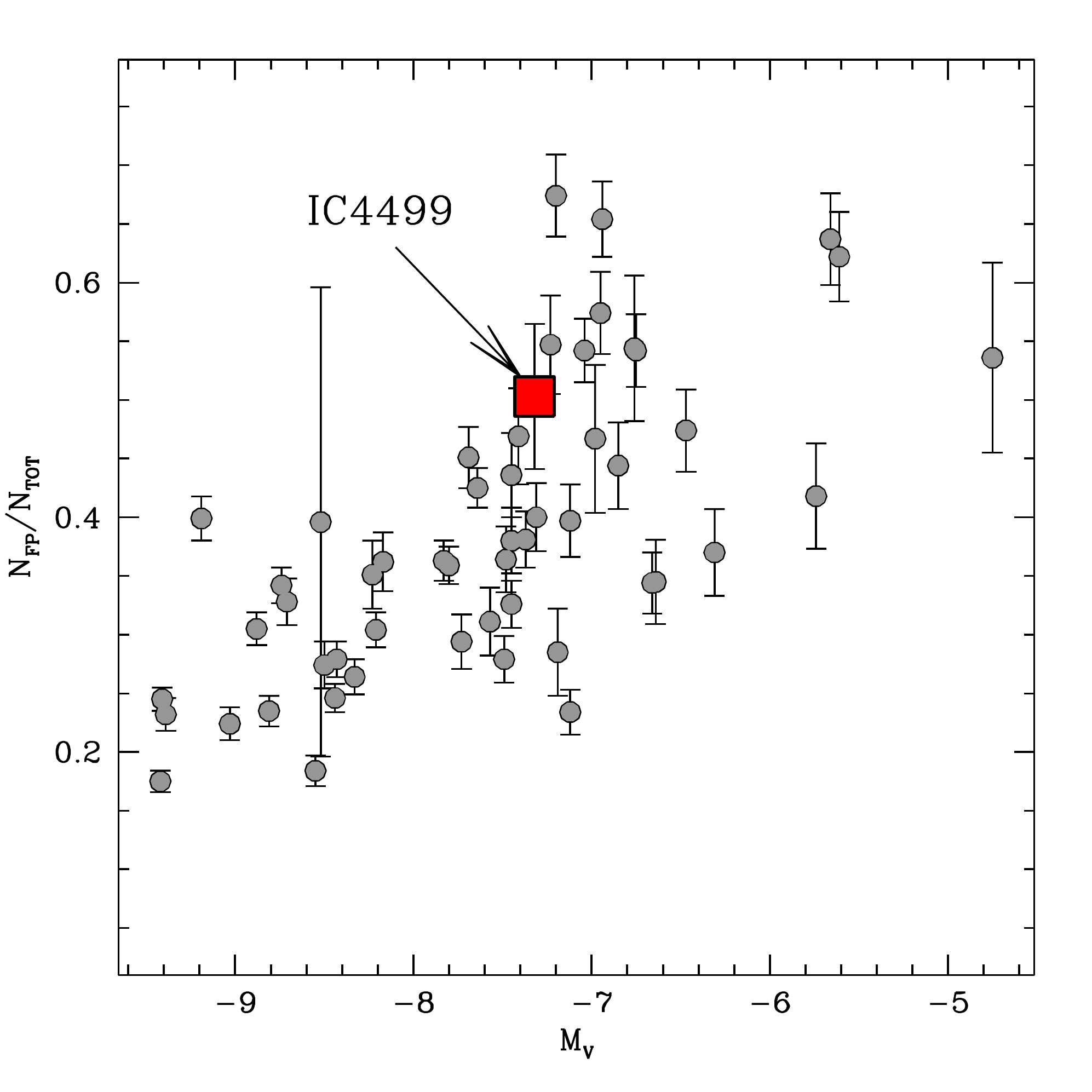}
	\caption{Fraction of FP stars respect to the total number of RGB stars (counted in the same magnitude range) 
	as a function of the 
	absolute cluster magnitudes. IC~4499 is highlighted with a red square and it 
	follows the general behaviour described by the other GCs observed within 
	the HST UV Legacy Survey \citep{piotto15,milone17}. \label{fig:ratio-mass}}
\end{figure}

\begin{figure*}
\begin{center}
\includegraphics[width=0.8\textwidth]{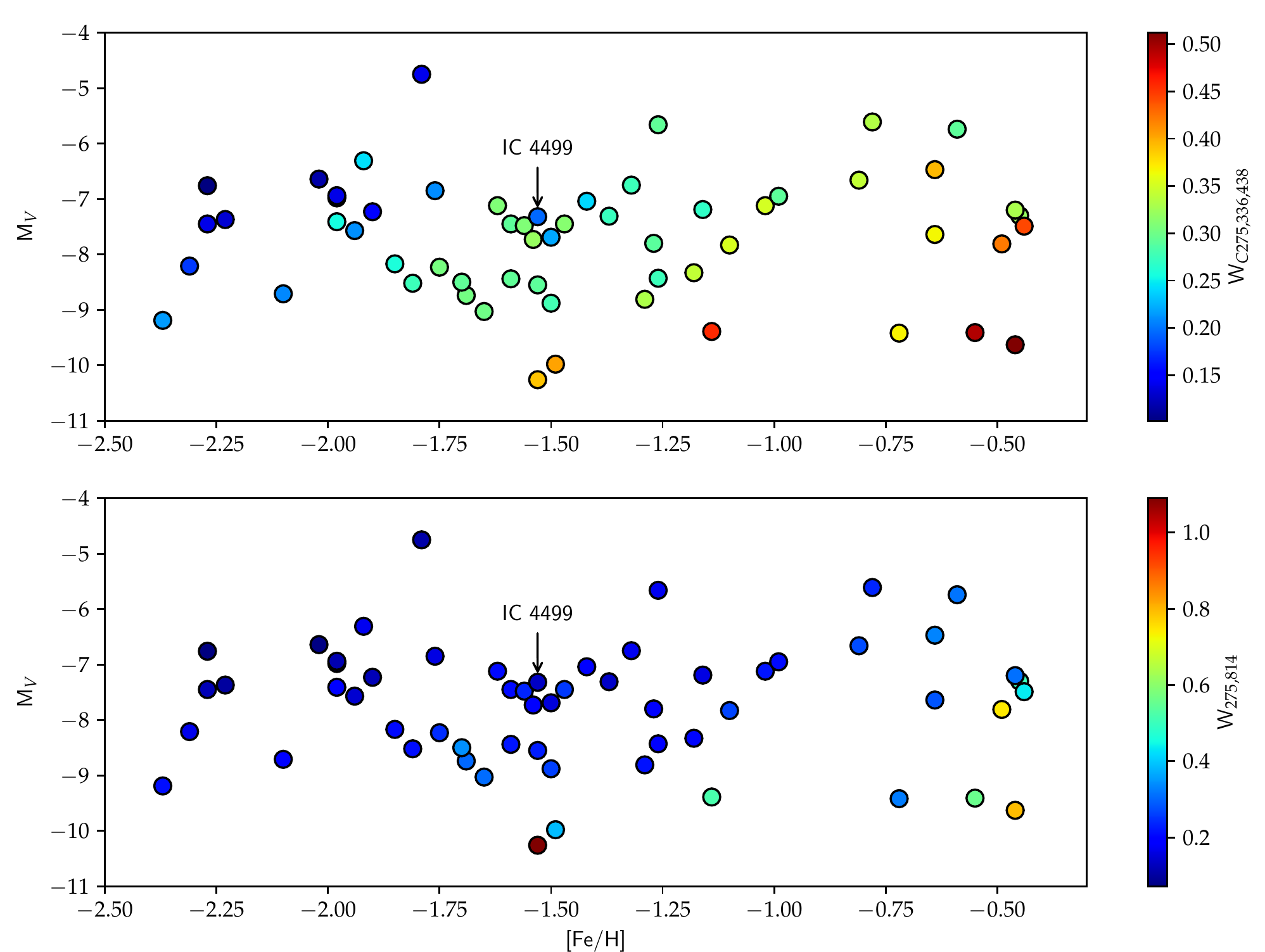}
	\caption{Metallicity vs. 
integrated $V$ magnitude distribution of GCs observed within 
	the HST UV Legacy Survey. Clusters are colour coded according to the values of $W_{C(275,336,438)}$ ({\em top panel}) and 
 $W_{275,814}$  ({\em bottom panel}). IC 4499 is highlighted in both plots.
  \label{fig:ic4499femass}}
\end{center}
\end{figure*}

\section{Results}
\subsection{Photometric evidence of MPs}
To test the presence of MPs in IC~4499 we first used the pseudo-colour diagram ($m_{F336W}$, 
$C_{275,336,438}$) shown in Figure~\ref{fig:cmap} 
($C_{275,336,438}=(m_{F275W}-m_{F336W})-(m_{F336W}-m_{F438W})$), as it traces 
simultaneously the OH, NH and CH molecular bands strength \citep[e.g.][]{piotto15,milone17}.
In this diagram we defined two fiducial lines (left panel of Figure~\ref{fig:cmap}) 
following the bluest and reddest end of the RGB and 
we then verticalised the distribution of RGB stars in the magnitude 
range $18.2<m_{F336W}<20.5$. 
The verticalised colour distribution ($\Delta_{C_{275,336,438}}$) is clearly bimodal with two peaks at 
$\Delta_{C_{275,336,438}}\sim0.03$ and $\sim0.14$ respectively (middle panel of Figure~\ref{fig:cmap}).
The detected bimodality is statistically significant, and a unimodal distribution 
can be rejected with a confidence level $>3\sigma$ 
according to a Gaussian Mixture Models analysis \citep{muratov10}.
{\it This result represents the first evidence of the presence of MPs in IC~4499.}

We classified the two sub-populations as FP and SP moving from red to blue colours 
as stars are expected to be progressively enhanced in N at decreasing $C_{275,336,438}$ colours.
From the areas under the Gaussian functions, we computed the number ratios between the sub-populations. 
We find that 97 stars can be attributed to the FP and 96 to the SP sub-sample, thus yielding 
$N_{FP}/N_{TOT}=0.503\pm0.062$,
where $N_{TOT}$ is the total number of RGB in the considered magnitude range. 
As shown in Figure~\ref{fig:ratio-mass}, given the absolute magnitude of IC~4499 
($M_V=-7.32$; \citealt{harris96}), the derived value of $N_{FP}/N_{TOT}$ 
appears to be compatible with the general ($N_{FP}/N_{TOT}$, $M_V$) trend observed 
for other GCs analysed adopting the same observational strategy (instrument and filters) as for
IC~4499.

\begin{figure}
\includegraphics[width=\columnwidth]{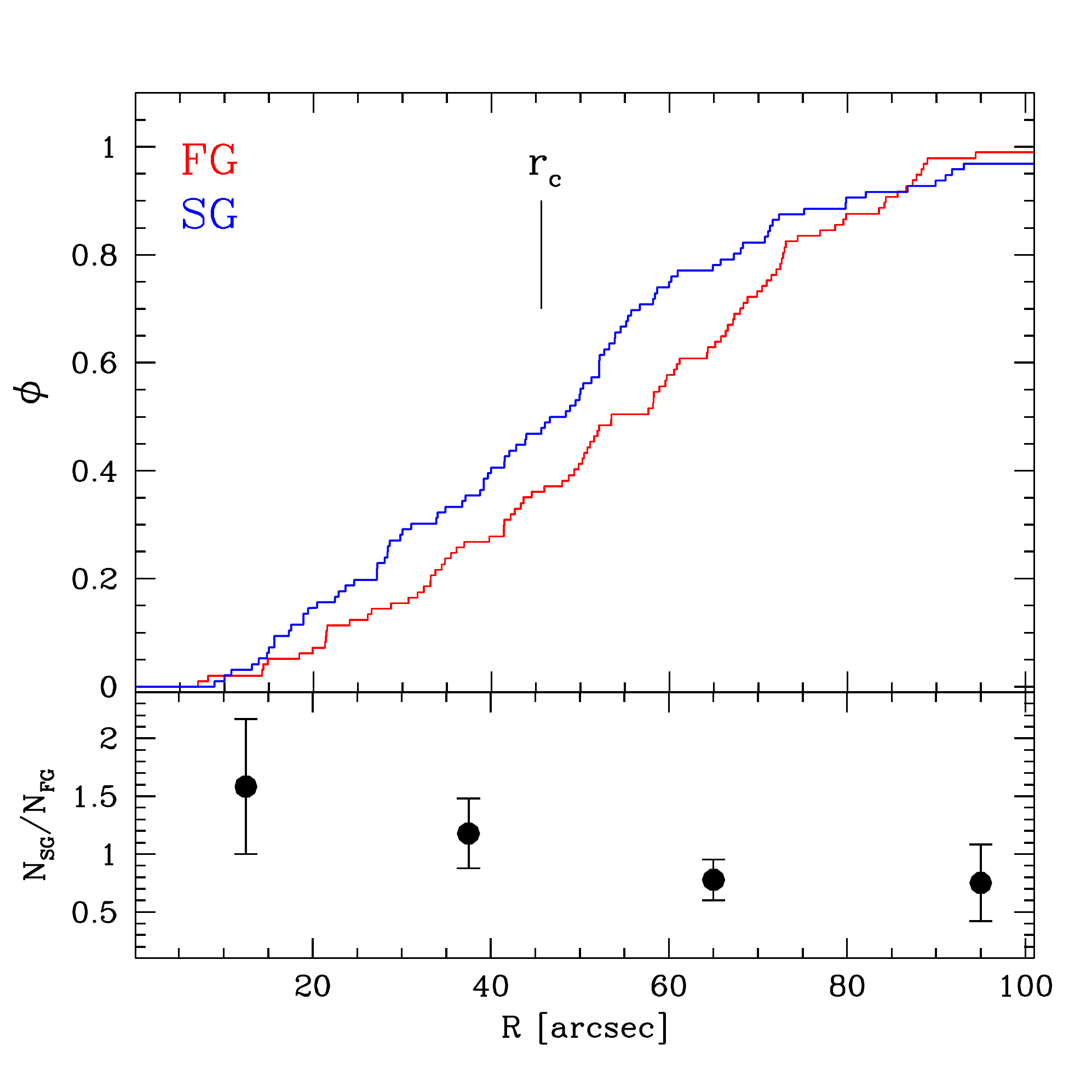}
	\caption{Upper panel: cumulative radial distribution of FP (red), and SP (blue) stars. 
	The lower panel shows the ratio between SP and FP stars as a function of the distance from the cluster centre.
	The position of the core radius ($r_c$) is also shown. \label{fig:ic4499_rad}}
\end{figure}

We have also computed the ($\Delta_{275,814}$, $\Delta_{275,336,438}$) diagram 
(the so-called ``chromosome map''; \citealp{milone17}), which
is shown in the right panel of Figure~\ref{fig:cmap}. $\Delta_{275,814}$ 
was obtained by verticalising the distribution of RGB stars in the
($m_{F814W}, m_{F275W}-m_{F814W}$) CMD in the same magnitude range used before. 
MPs are clearly separated also in this diagram. We also note 
that both FP and SP stars are
well clumped in the ``chromosome map'' attaining very similar ($m_{F275W}-m_{F814W}$) colours thus 
suggesting a very
small (if any) He variation within each and among the two sub-populations \citep[e.g.][]{milone15,lardo18}.
This result is further confirmed by the relative colour distribution of FP and SP stars 
in the ($m_{F606W}$, $m_{F606W}-m_{F814W}$)
CMD. In fact, we find that their average colour difference 
is well within the photometric errors along the RGB.
For comparison, in Figure~\ref{fig:cmap} we also show the 
($\Delta_{275,814}$, $\Delta_{275,336,438}$) distribution for NGC~288, which is a well studied low-mass 
($M_V=-6.75$; \citealt{harris96}) cluster with similar metallicity to IC~4499.

The intrinsic RGB widths both in ($\Delta_{275,814}$) and ($\Delta_{275,336,438}$) colour combinations
were derived for stars 2 magnitudes brighter than the TO in the $m_{F814W}$ 
band. We subtracted photometric errors and potential residuals of the differential reddening corrections 
to the observed widths and we obtain $W_{C(275,336,438)}=0.195\pm0.009$ and $W_{275,814}=0.132\pm0.008$. 
Interestingly, Figure ~\ref{fig:ic4499femass} shows that the RGB width $W_{C(275,336,438)}$ of IC~4499 
is significantly smaller than what observed in other 
clusters with similar metallicity and mass, thus suggesting a significantly smaller enhancement in N abundance.
Only one cluster, namely NGC~6584, has a similar RGB width ($W_{C(275,336,438)}=0.195\pm0.009$; \citealt{milone17}) 
in the same metallicity and mass range.

We have also derived the radial distribution of FP and SP sub-populations.
To this aim, we first calculated the cluster centre ($C_{grav}$) following the method reported by 
\citet[][see also \citealt{cadelano17}]{dalessandro18}, which is based on direct star counts.
Briefly, $C_{grav}$ has been determined following an iterative procedure that
starts from a first-guess centre and computes the average of the coordinates 
($\alpha$ and $\delta$) of a sub-sample of stars located within a circle of radius $r$ and 
in a defined magnitude interval. 
 The procedure stops when convergence is reached, i.e., when the newly determined centre coincides 
with the previous ones within an adopted tolerance limit. 
We repeated the procedure 27 times, using different values of $r$ and selecting stars in 
different magnitude ranges, chosen as a compromise between having high enough statistics and avoiding spurious 
effects due to incompleteness and saturation. 
In particular, the radius $r$ has been chosen in the range of $85\arcsec-100\arcsec$ with a step of $5\arcsec$, 
thus guaranteeing that it is
always larger than the literature core radius $r_c = 50.4$ \citep{harris96}. 
For each radius $r$, we have explored six magnitude
ranges, from $m_{F606W} \geq 14$ (in order to exclude stars close to the
saturation limit), down to $m_{F606W}=22$ , in steps of $0.3-0.2$ mag. 
As the first-guess centre, we used the one quoted by \citet{shawl86}. 
The final value adopted as $C_{grav}$ is the mean of the different 
values of $\alpha$ and $\delta$ obtained in the 27
explorations, and its uncertainty is their standard deviation.
The derived $C_{grav}$ results to be located at $\alpha = 15^h0^m18.749^s$ and $\delta =-82^{\circ}12\arcmin52.486\arcsec$, 
with a total uncertainty of about $2\arcsec$. 
This newly determined center is $3.2\arcsec$ distant from the one quoted by \citet{shawl86}.

The cumulative radial distributions of FP and SP stars with respect to $C_{grav}$ and 
the number ratio $N_{SP}/N_{FP}$ radial variations within the entire WFC3 FOV are shown in
Figure~\ref{fig:ic4499_rad}. SP stars are more centrally concentrated than FP. 
The Kolmogorov-Smirnov test gives a probability $P_{KS}\sim0.03$ that the two populations are 
extracted from the same parent population. 
However, it is important to stress that the HST data cover only $\sim 1 r_h$ ($r_h\sim 102\arcsec$; 
\citealt{harris96}). 
As discussed in
\citet{dalessandro18}, radial distributions obtained over a limited area 
($<1.5-2\times r_h$) might not be representative of the overall relative distribution and therefore 
should be taken with some caution.

 \begin{figure}
\includegraphics[width=\columnwidth]{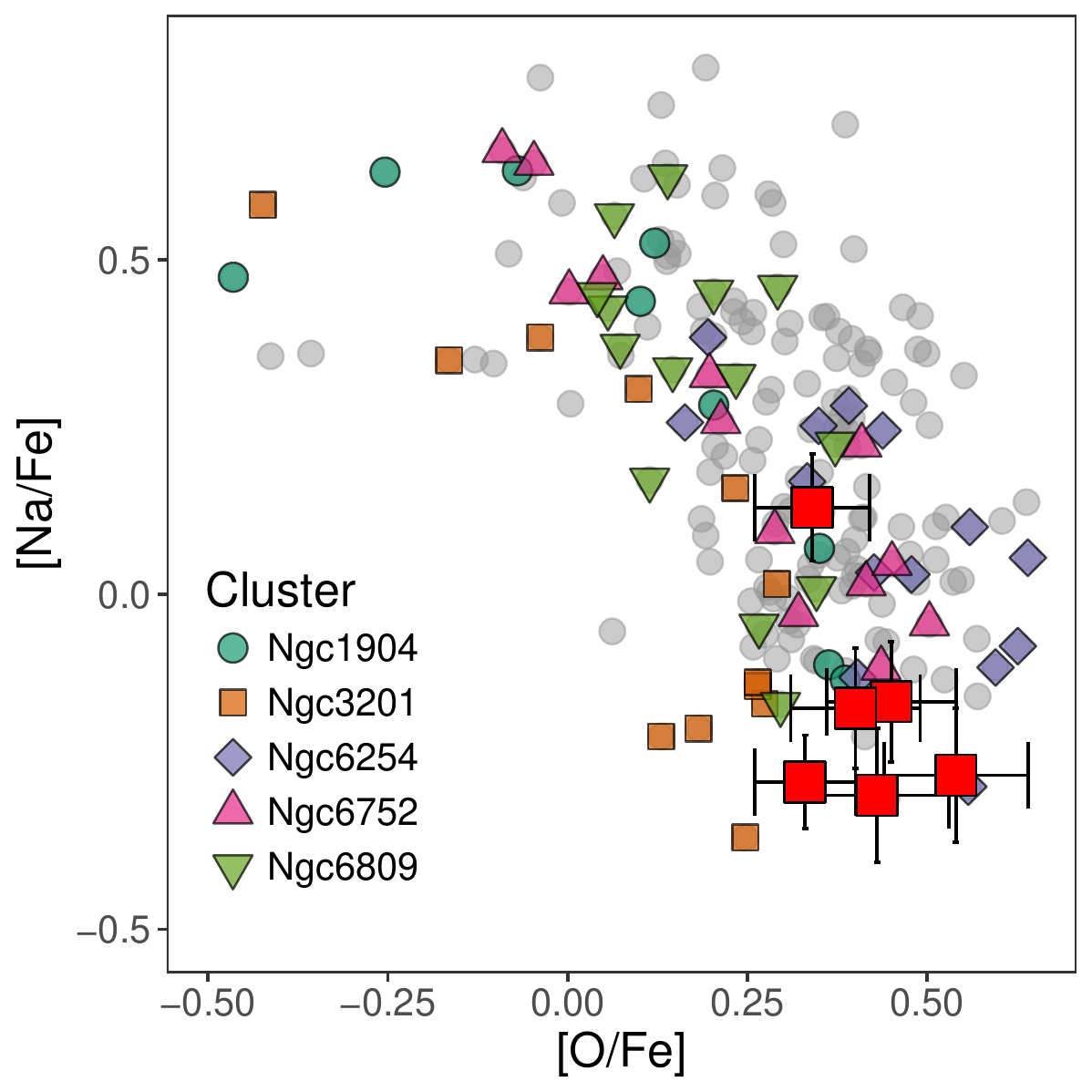}
\caption{IC4499 [Na/Fe] and [O/Fe] abundance ratios (red squares) are over-plotted to the homogeneous 
data from the \citet{carretta2009} survey of ancient Galactic globulars (grey circles). 
Clusters with similar metallicity as IC~4499 are also highlighted (see legend).}
        \label{NAO}
   \end{figure}

\subsection{Abundance results}
The chemical analysis of IC~4499 provides a mean metallicity of [FeI/H] = --1.64 ($\sigma$=0.05) dex and 
 [FeII/H] = --1.57 ($\sigma$=0.04) dex. This value is slightly lower (but still consistent) than previous
derivations by \citet{hankey11} who measured  [Fe/
H] = --1.52 $\pm$ 0.12 dex. However, these
analyses were based on low-resolution spectra
around the calcium triplet feature that is more prone to uncertainties and to
systematics related to the adopted calibrations.

The mean [O/Fe] = 0.42 ($\sigma$=0.07) dex, with no indication of intrinsic spreads.
Five out of six stars have also very similar Na abundances within the uncertainties. 
On the other hand, one star (S634 - see Table~2) has higher Na abundance by $\sim0.4$dex, 
whereas both O and Mg abundances are consistent with the other stars in the sample.  
This behaviour is consistent with the Na-O anti-correlation observed in other GCs
as shown in Figure~\ref{NAO} and obtained by \citet{carretta2009}. While in GCs with global extended Na-O anti-correlation, 
both [O/Fe] and [Na/Fe] display large ranges, for small [Na/Fe] variations around the field star area,
[O/Fe] stay almost constant, as for the case of IC~4499 (red squares in Figure~\ref{NAO}).
This result further confirms the evidence that IC~4499 does host MPs.
The detection of only one Na enriched star (corresponding to $16\pm18$\% of the sample) does not match the expectations 
based on photometric results of almost equi-populated sub-populations. However, it should be noted that spectroscopic targets 
are distributed well outside the HST FOV, where the SP/FP ratio appears to be significantly lower than in the central regions.

Consistently with what constrained by using UV photometry, 
IC~4499 shows the least extended 
Na-O anticorrelation compared to clusters with similar metallicity. 
However, the significance of such a comparison is limited by the low-number statistics of our sample. 

For Mg, we find a mean abundance of  [Mg/Fe] = 0.30 ($\sigma$=0.07) that well compares with the mean [Mg/Fe] 
abundance ratio measured by 
\citet{carretta2009} in NGC~6254, NGC~6752, and NGC~6809. 
As for O, Mg does not show intrinsic abundance variations.

\section{Summary and conclusions}
IC~4499 was suggested 
to be one of the very few old Galactic globular clusters to not show MPs \citep{walker11}. 
This result was based on multi-band ground-based photometry, 
targeting a limited number of stars and likely dominated by the FP sub-populations as observations were mainly 
biased towards the external regions of the cluster.
In this study, we present the first evidence that IC~4499 does host MPs with light-element abundance variations.

Our analysis is based on both high-resolution photometric and spectroscopic data. 
The UV and optical HST photometry shows that the RGB of IC~4499 clearly splits in two sequences, which are also
well identified in the so called ``chromosome map'' (($\Delta _{275,814}$), ($\Delta _{C 275,336,438}$) diagram)
thus suggesting they have different light-element abundances. 
This result is qualitatively confirmed by the Na abundance distribution.  
In fact, we find that one star has Na abundance compatible with being SP as it is enriched by $\sim 0.4$ dex.
Conversely both O and Mg do not show any evidence for internal variations.

Interestingly, we find that the fraction of SP with respect to FP stars well fits the trend 
with the present-day cluster mass observed for other Galactic GCs \citep[e.g.][]{milone17}.
On the contrary, the width of the RGB in both ($\Delta _{275,814}$)  and ($\Delta _{C 275,336,438}$) -- which are proxies for 
the Y and N enhancement, respectively \citep[e.g.][]{milone15,lardo18} -- are significantly smaller 
than the values derived for old Galactic globulars with similar metallicity and mass by \citet{milone17}. 
A similar indication is also suggested by the Na abundance distribution, although it is not conclusive because of 
the small sample size.

\citet{hankey11} suggested a possible extra-Galactic origin of IC~4499. In fact, based on the mean 
radial velocity of the system, the authors found a possible connection with the Monoceros ring, 
which would eventually suggest that
IC~4499 formed in the disrupted Canis Major dwarf galaxy \citep{helmi03,martin05}.
In this context, it would be tempting to speculatively connect 
the moderate N and Na abundance variations observed in IC~4499 with its extra-Galactic origin.
GCs formed in the Sagittarius dwarf galaxy and Fornax typically show small variations in terms 
of Na and O (with the obvious exception of the very massive system M~54; \citealt{carretta14}).  
In addition, they tend to show less extended horizontal branch morphologies (so possibly smaller He spreads)
and redder far-UV indexes than genuine Milky-Way GCs with similar mass and metallicity \citep{dalessandro12}. 

While this trend is surely fascinating and possibly suggestive of a link between the environment and the efficiency 
of light-element enrichment a cluster can undergo, at the present stage it is hard to firmly disentangle
it from the contribution of other key parameters, such as cluster mass and age.
Moreover, a robust distinction between genuine Milky-Way and accreted systems is of crucial importance 
in this context. The exquisite astrometric performance of Gaia will definitely help to firmly constrain 
the extra-Galactic origin of some GCs and to unveil possible new accreted systems 
(see for example \citealt{myeong18}), thus possibly shedding new light 
on the actual role played by the environment on the properties of MPs.

\section*{Acknowledgements}
The authors thank the referee for the careful reading of the paper.
ED acknowledges financial support from the Leverhulme Trust Visiting Professorship Programme VP2-2017-030.
ED is also grateful for the warm hospitality of LJMU where part of this work was performed.
CL acknowledges financial support from the Swiss National Science Foundation (Ambizione grant PZ00P2\_168065).
NB gratefully acknowledges funding from the ERC under the European Union's Horizon 2020 research and innovation programme via 
the ERC Consolidator Grant Multi-Pop (grant agreement number 646928, PI Bastian). NB is a Royal Society University Research Fellow.





\bibliographystyle{aa}







\end{document}